\documentclass[10pt,preprint2]{aastex} % for preprints % for astro-ph

\shorttitle{X-ray Sources in M100}
\shortauthors{Kaaret}

\begin{document}

\title{Chandra X-Ray Point Sources, including Supernova
1979C, in the Spiral Galaxy M100} 

\author{Philip Kaaret}

\affil{Harvard-Smithsonian Center for Astrophysics, 60 Garden
St., Cambridge, MA 02138, USA}

\email{pkaaret@cfa.harvard.edu}

\begin{abstract}

Six x-ray point sources, with luminosities of $4 \times
10^{38} - 2 \times 10^{39} \rm \, ergs \, s^{-1}$ in the
0.4--7~keV band, were detected in Chandra observations of the
spiral galaxy M100.  One source is identified with supernova
SN 1979C and appears to have roughly constant x-ray flux for
the period 16--20 years after the outburst.  The x-ray
spectrum is soft, as would be expected if the x-ray emission
is due to the interaction of supernova ejecta with
circumstellar matter.  Most of the other sources are variable
either within the Chandra observation or when compared to
archival data.  None are coincident with the peak of the
radio emission at the nucleus.  These sources have harder
spectra than the supernova and are likely x-ray binaries.
M100 has more bright x-ray sources than typical for spiral
galaxies of its size.  This is likely related to active star
formation occurring in the galaxy.

\end{abstract}


\section{Introduction}

M100 (NGC 4321) is a `grand-design' late-type spiral, which
has hosted 4 supernovae since 1900, has a prominent,
star-forming circumnuclear region, and is located in the
Virgo cluster . Previous x-ray observations of M100 led to
the detection of the most recent supernova in the galaxy (SN
1979C) and of a number of other point sources
\citep{immler98,palumbo81}. 

Here, we report on x-ray point sources found in high-angular
resolution x-ray observations of M100 made with Chandra. 
Results on the nuclear emission from the same data were
previously reported in \citet{ho01}.  We selected this galaxy
for study because it is seen nearly face-on which reduces
ambiguity in determining the environments of x-ray sources,
because it contains SN 1979C which was previously detected in
x-rays \citep{immler98}, and because of the question of the
association of an x-ray source with the nucleus remains open
\citep{ho01}.  We describe the Chandra observations and our
analysis in \S 2 and discuss the results in \S 3.

\section{Observations and Analysis}

M100 was observed with the Chandra X-Ray Observatory (CXO;
Weisskopf 1988) on 6 Nov 1999 and 15 Jan 2000 using the ACIS
spectroscopic array (ACIS-S; Bautz et al.\ 1998) in imaging
mode and the High-Resolution Mirror Assembly (HRMA; van
Speybroeck et al.\ 1997). The galaxy fits almost completely
on the S3 chip with a small portion of the $D_{25}$ ellipse
on the S2 chip.  We found no point sources within the portion
of the $D_{25}$ ellipse on the S2 chip and the analysis below
includes only data from S3.  Level 2 data products from the
standard processing (version R4CU5UPD11.2) were used.  The
processing includes standard cuts on CCD event grades,
elimination of events from hot pixels, hot columns, and near
node boundaries, and removal of cosmic ray induced events.

\begin{deluxetable}{lccccccccc}
\tablecolumns{10}
%\tabletypesize{\scriptsize}
\tablecaption{Chandra X-Ray Point Sources in
M100.\label{chandra_src}}
\tablewidth{0pt}
\tablehead{\colhead{Source} & \colhead{RA} & \colhead{DEC} &
   \colhead{$\log(L_{X})$} & \colhead{Color} & \colhead{Obs 1}
   & \colhead{Obs 2} }
\startdata
C1 & 12 22 54.78 & +15 49 18.3  & 39.1(1) & -0.2(2) & 39.3(1) & 38.6(2) \\
C2 & 12 22 54.13 & +15 49 11.9  & 39.1(1) & -0.1(2) & 38.9(2) & 39.2(1) \\
C3 & 12 22 54.76 & +15 49 15.9  & 39.0(1) &  0.4(3) & 39.0(1) & 39.0(1) \\
C4 & 12 22 54.23 & +15 49 43.8  & 38.8(1) &  0.2(3) & 38.6(2) & 38.9(2) \\
C5 & 12 22 58.66 & +15 47 51.0  & 38.9(1) & -0.5(3) & 38.9(2) & 38.8(2) \\
C6 & 12 22 46.10 & +15 48 49.7  & 38.9(1) &  0.6(3) & 39.1(1) & 38.5(3) \\
\enddata

\tablecomments{This Table contains for each source: Source --
the label used in this paper; RA and DEC -- the J2000
position; $\log(L_{X})$ -- the logarithm of the luminosity in
the 0.4--7~keV band in units of $\rm ergs \, s^{-1}$; Color
-- an x-ray color (hardness ratio) defined as counts in the
1-7~keV band minus counts in the 0.4--1~keV band divided by
the total counts in the 0.4--7~keV band; Obs 1 and Obs 2 --
logarithms of the luminosities in the first and second
Chandra observations.  Numbers in parentheses indicate the
$1\sigma$ uncertainty in the final digit of the preceding
value.}  \end{deluxetable}

\begin{figure*}[th] \epsscale{1.6} \plotone{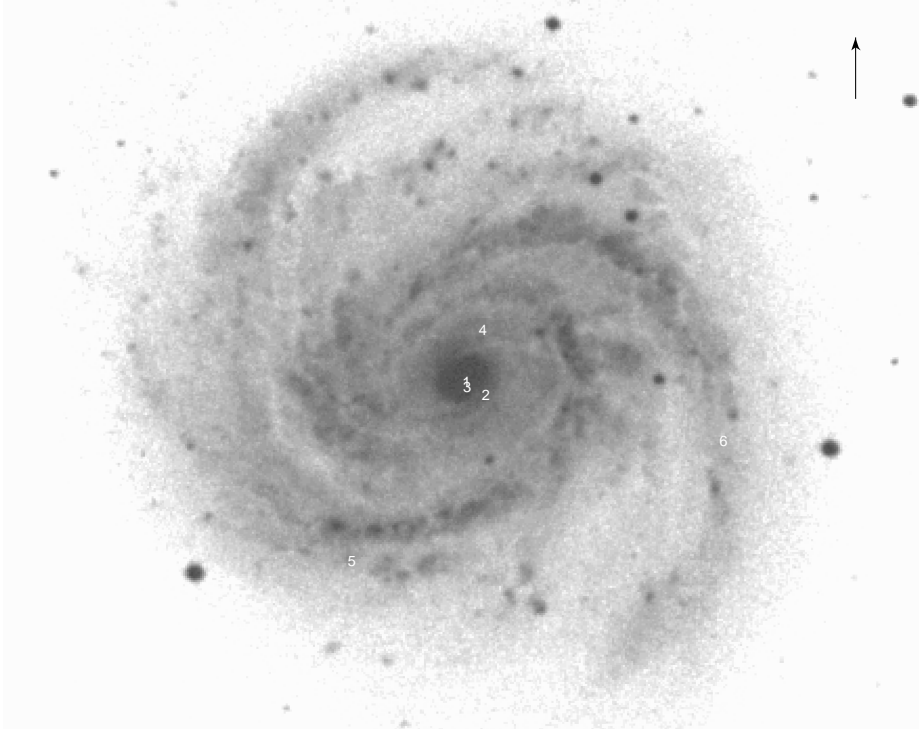}
\caption{B-band image of M100 with positions of the Chandra
point sources superimposed.  The numbers indicate x-ray
sources from Table~\ref{chandra_src}.  The arrow points
North.  The image is from the Digitized Sky Survey.}
\label{dssb_x} \end{figure*}

The observation on 6 Nov 1999 lasted 1165~s and had a high
background rate in S3, $10-70 \rm \, cts \, s^{-1}$. On 15
Jan 2000, 1366~s of data were obtained with a low background
rate, $\sim 1 \rm \, cts \, s^{-1}$.  Both observations are
recorded in Chandra ObsID 400.  We created images in the
0.4--7~keV band at the full ACIS resolution of $0.49\arcsec$
for each observation and for the combined data.  We searched
the images for point sources using wavdetect (CIAO V2.0) and
visually inspected each image.  For the two sources near the
nucleus of M100, wavdetect reported source ellipses well in
excess of the HRMA/ACIS point spread function.  These large
source ellipses are likely due to diffuse emission in this
region and we chose to use source regions consistent with the
point spread function rather than those reported by
wavdetect.

The merged list of sources with detection significance above
$4\sigma$ and within the $D_{25}$ ellipse of M100 is given in
Table~\ref{chandra_src}.  The positional accuracy is limited
by the Chandra absolute aspect which is accurate to
$0.6\arcsec$ ($1\sigma$ rms; Aldcroft et al.\ 2000; see also
the Chandra Aspect web pages
http://asc.harvard.edu/mta/ASPECT/).  The relative positions
should be somewhat better.  For each source, counts from the
full data set in the energy bands 0.4--1--7~keV were
extracted from source regions with a radius of 2 pixels for
sources near the aimpoint and larger elliptical regions for 
sources further out.  The counts were background subtracted
using a circular background regions with radii equal to 3
times the major axis of the source ellipse and with ellipses
twice the size of the source regions excluded (for both the
source of interest and any sources overlapping the background
region).   For each source, we calculated the x-ray count
rate in the 0.4--7~keV band and an x-ray color (hardness
ratio) defined as counts in the 1-7~keV band minus counts in
the 0.4--1~keV divided by the total counts in the 0.4--7~keV
band.  The x-ray colors were corrected to the values which
would have been measured at the aimpoint using exposure maps
for 0.7 and 2.4~keV, which correspond to the average photon
energies for the point sources in the two bands.

Count rates were converted to luminosities in the 0.4--7~keV
band assuming a thermal bremsstrahlung model with $kT = 2 \rm
\, keV$ and corrected for absorption along the line of sight
with a column density of $2.39 \times 10^{20} \rm \, cm^{-2}$
and assuming a distance of 16.1~Mpc \citep{ferrarese96}.  The
count rates were corrected for the fraction of events
contained in the extraction region and for exposure using an
exposure map for 1.5~keV.  The $kT = 2 \rm \, keV$ thermal
bremsstrahlung spectrum is consistent with the x-ray colors
for most of the sources (see below).  The previously used $kT
= 5 \rm \, keV$ thermal bremsstrahlung spectrum
\citep{immler98} is not consistent with the x-ray colors. 
However, use of $kT = 5 \rm \, keV$ would increase the
luminosities only by a factor of 1.3.  Use of a power-law
with photon index 2 would increase them by 1.1.  For
comparison with Rosat measurements \citep{immler98}, the
luminosities in Table~\ref{chandra_src} should be multiplied
by 1.0 to find the equivalent luminosity in the 0.1--2.4 band
assuming a thermal bremsstrahlung model with $kT = 5 \rm \,
keV$ for a distance of 17.1~Mpc.

\section{Results and Discussion}

\begin{figure*}[tb] \epsscale{1.4} \plotone{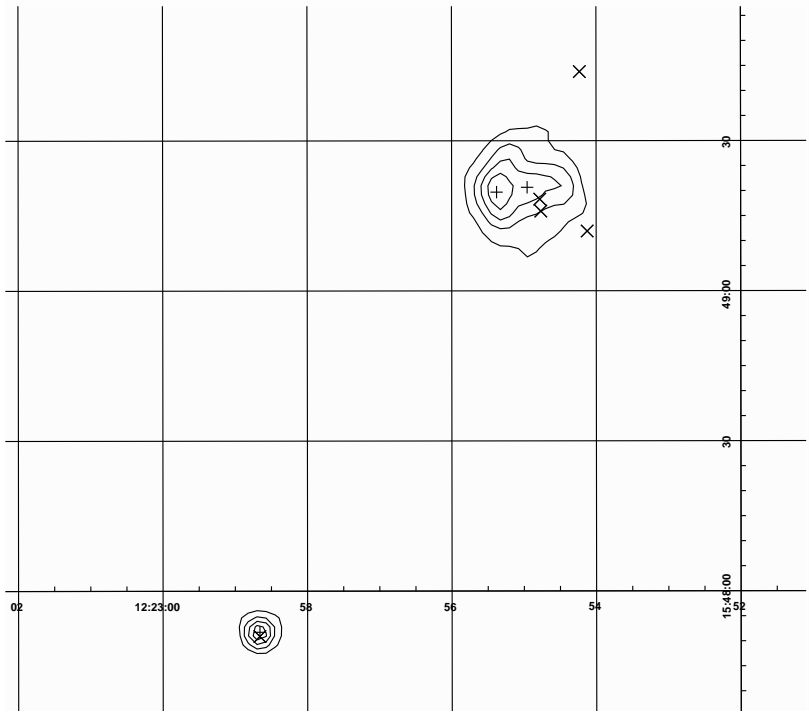}
\caption{Contour map of 1.4~GHz radio emission from M100. 
The contours are at 1, 2, 3, 4 mJy/beam.  Crosses indicate
positions of sources from the FIRST catalog and X's indicate
x-ray sources from Table~\ref{chandra_src}.  The radio
emission towards the upper-right is near the nucleus while
the emission towards the lower-left is from SN 1979C.}
\label{firstmap} \end{figure*}

Table~\ref{chandra_src} gives the properties of the x-ray
point sources detected with Chandra.  Fig.~\ref{dssb_x} shows
the locations of the Chandra sources superimposed on a
Digitized Sky Survey Blue image of M100.  The Chandra sources
lie either on spiral arms, in H{\sc II} regions near spirals
arms, or near the nucleus, i.e.\ at regions of active star
formation.  This supports the idea that luminous x-ray
sources ($L_{X} > 2 \times 10^{38} \rm \, ergs \, s^{-1}$)
are preferentially associated with recent star formation
\citep{fabbiano89,fabbiano01}.

Comparing the sources detected with Chandra with those
detected with Rosat and Einstein \citep{immler98}, we find
several variable sources.  Allowing for the uncertainty in
the luminosity due to the lack of spectral information, we
require a factor of $\sim 2$ change as evidence of
variability.  The Chandra source C6 was undetected with Rosat
and appears to have increased in flux by a factor of at least
3.  The Chandra upper bound on the luminosity of the Rosat
source H17 \citep{immler98} is at least a factor of 4 below
the Rosat measurement.  The Rosat source H24 appears to have
decreased in flux by a factor of at least 2 in the Chandra
observation.  The source C2 = H21 may also be variable, but
caution is warranted before reaching this conclusion as the
Rosat flux may include some diffuse emission due to the
larger point spread function of Rosat. This source was also
detected with Einstein at a comparable flux.  The Rosat
sources H18 and H19 were detected at lower fluxes (the Rosat
observation was 43~ks compared to only 2.5~ks for Chandra)
and their variability cannot be constrained from the short
Chandra observation.  A longer Chandra observation would be
useful.  H19 was detected with Einstein at a much higher flux
than with Rosat and is clearly variable.  The sources H22 =
C4 and H25 = C5 appear constant.

The nucleus of M100 (Rosat source H 23) resolves with Chandra
into two sources: C1 and C3.  There is also diffuse emission
present in the low background image near the two point
sources.  C1 lies $2.7\arcsec$ away from the nuclear position
derived from Digitized Sky Survey images which has an
uncertainty of $1.7\arcsec$ \citep{cotton99}.  To further
investigate association of the C1 with the nucleus, we
examined the FIRST radio survey data at 1.4~GHz for M100
\citep{becker95}.  The FIRST catalog \citep{white97} contains
two sources inside a region of extended emission near the
nucleus, see Fig.~\ref{firstmap}.  C1 is not coincident with
either FIRST source and lies outside the peak of the diffuse
emission.  C1 lies $3.5\arcsec$ away from the closer FIRST
source (which lies within the optical nuclear error box). 
The offset between C1 and the FIRST source is significantly
larger than the astrometric uncertainty of Chandra of
$1.0\arcsec$ at 90\% confidence and the uncertainty of the
FIRST position which is less than $0.5\arcsec$ at 90\%
confidence.  The FIRST and Chandra positions for SN 1979C,
see below, agree within $0.84\arcsec$ indicating there are no
gross errors in the x-ray positions.  Hence, we exclude an
identification of C1 with the bright radio sources near the
nucleus of M100.  We note that M100 does not show evidence
for AGN activity at other wavelengths \citep{sakamoto95}.

C1 is the only source which shows significant variability
between the two Chandra observations.  The variability
indicates the source is likely a compact object.  The
circumnuclear region of M100 is undergoing vigorous star
formation with some star clusters as young as a few 10~Myr
\citep{ryder99}.  Hence, the presence of bright x-ray point
sources near the nucleus is not unexpected and is likely
related to the young stellar population.

\begin{figure}[tb] \epsscale{0.8} \plotone{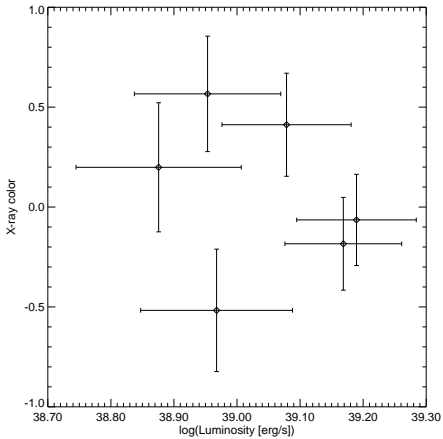}
\caption{X-ray color versus luminosity for point sources in
M100.} \label{color_lum} \end{figure}

The majority of the x-ray sources in M100 are variable and
are likely x-ray binaries.  The sources are not more luminous
than known x-ray binaries within the Milky Way, e.g.\
4U\,1543--47 \citep{orosz98}.  Hence, they may be normal
stellar black hole candidates x-ray binaries with black
masses near $10 \, M_{\odot}$ and, perhaps, mildly beamed. 
The luminosities are not high enough to warrant the
application of exotic models as may be required for very high
luminosity sources, e.g.\ the $L_{X} \sim 10^{41} \rm \, ergs
\, s^{-1}$ source in M82 \citep{kaaret01,matsumoto01}.  In
Fig.~\ref{color_lum}, we show the x-ray color versus x-ray
luminosity (both defined in Table~\ref{chandra_src}) of the
Chandra sources.  Excluding the softest source, which is a
supernova discussed below, the colors lie in the range from
-0.2, corresponding to a thermal bremsstrahlung spectrum with
$kT \sim 1.0 \rm \, keV$ or a power-law spectrum with photon
index $\alpha = 2.3$, to +0.6, corresponding to $kT \sim 2.3
\rm \, keV$ or $\alpha = 0.6$, assuming a column density of
$2.39 \times 10^{20} \rm \, cm^{-2}$.  Such spectra are
reasonable for accreting x-ray binaries at these
luminosities.  

Using the x-ray point source luminosity function derived from
a Rosat survey of nearby spiral galaxies \citep{roberts00},
we estimate that roughly 1.4 x-ray sources with $L_X \ge 5
\times 10^{38} \rm ergs \, s^{-1}$ would be expected in M100
given its blue luminosity.  We find 6 sources.  The number of
x-ray point sources versus blue luminosity in M100 is similar
to that found for M101 \citep{wang99} and is likely related
to active star formation, as has been suggested for M101
\citep{trinchieri90}.  Star formation in M100 is concentrated
along the spiral arms \citep{knapen96} as are the Chandra
sources.

\subsection{SN 1979C}

The Rosat source H25 was identified with the supernova SN
1979C \citep{immler98}.  Our Chandra position places the
source C5 only $0.84\arcsec$ away from the supernovae --
within the 90\% confidence error circle -- and greatly
strengthens this association.  None of the other known
supernovae are detected with Chandra.  SN 1901B and SN 1959E
are type I supernova \citep{barbon89}, so no x-ray emission
would be expected \citep{schlegel95}.  The type of SN 1914A
is not known.

Unfortunately, there are too few photons ($\sim 10$) detected
with Chandra from SN 1979C to make definitive statements
about the spectrum (e.g.\ whether it is thermal line emission
versus continuum emission).  The source is the softest of
those found in M100.  The x-ray color, defined above, of
$-0.5 \pm 0.3$ is consistent with a thermal bremsstrahlung
spectrum with $kT = 0.3-1.1 \rm \, keV$, or a power-law
spectrum with photon index of 2.3 to 4.3.  This assumes a
column density of $2.39 \times 10^{20} \rm \, cm^{-2}$;
higher absorption would imply a softer intrinsic spectrum. 
The spectrum is significantly softer than the assumed
spectrum used in calculating luminosities in
Table~\ref{chandra_src} or in \citet{immler98}. 
Recalculating the luminosity using a more appropriate
spectrum, thermal bremsstrahlung with $kT = 0.6 \rm \, keV$,
we find a luminosity of $9 \times 10^{38} \rm erg \, s^{-1}$
in the 0.1--2.4~keV band and a luminosity from the Rosat data
of $1.3 \times 10^{39} \rm erg \, s^{-1}$ in the same band.  

The Chandra and Rosat fluxes for the source H25 = C5 are not
strongly inconsistent given the uncertainties due to spectral
shape and inter instrument calibration.  The source appears
to have had a roughly constant, or perhaps slightly
declining, luminosity over the 4.3 years separating the two
observations.  The constancy of the x-ray luminosity of SN
1979C over 4~years beginning 16 years after outburst is
similar to the x-ray light curve of SN 1978K in NGC 1313
which showed a constant luminosity of $\sim 5 \times 10^{39}
\rm erg \, s^{-1}$ over the period 12--16 years after
outburst \citep{schlegel96}.

SN 1979C was discovered on 19 April 1979 near maximum optical
light \citep{johnson79} and has been extensively observed. 
The optical light curve had a linear decline, leading to
classification as a type II, subclass `L' or `linear'
\citep{panagia80}.  The total radiative energy in the event
was $7 \times 10^{49} \rm erg$ \citep{panagia80}.  The
progenitor is thought to be a red supergiant with a mass near
$17 \, M_{\odot}$ and may have been in a binary
\citep{weiler92,schwarz96}.  Hubble Space Telescope (HST)
observations led to tentative identification of an optical
counterpart to SN 1979C \citep{vandyk99}.  The H$\alpha$ flux
from the counterpart would imply an H$\alpha$ to X-ray flux
ratio of $\sim 0.013$ during 1995--1996.

The interaction of supernova ejecta with surrounding matter
previously lost by the progenitor provides an adequate
explanation of optical, UV, and radio data on SN 1979C
\citep{chevalier94}.  Radio emission was detected one year
after outburst with the maximum emission at 6~cm shortly
after the first detection and the maximum at 20~cm three
years after outburst \citep{weiler81,weiler86}.  This delayed
turn-on with increasing delay at longer wavelengths appears
due to decreasing free-free absorption in ionized gas around
the supernova and is well explained by the circumstellar
interaction model \citep{weiler91}.  The same model also
explains the observed UV lines \citep{fransson84}.  

Circumstellar interactions could produce the observed roughly
constant luminosity if the supernova ejecta and the
circumstellar matter have the appropriate radial dependence
\citep{schlegel96}.  Such a process could produce adequate
luminosity and the H$\alpha$ to X-ray flux ratio is in
reasonable agreement with model predictions
\citep{chevalier94}.  A soft x-ray spectrum would be
expected, which is consistent with the Chandra constraints.
The circumstellar interaction model appears consistent with
all the available x-ray constraints.

Alternative explanations of the x-ray emission are possible
if a compact object was formed in SN 1979C.  The X-ray
emission might arise from a rotation-powered pulsar or from
accretion on to a compact object fueled either by a binary
companion or by matter from the progenitor which has fallen
back onto the compact object \citep{chatterjee00}.  
Sufficient flux could be produced by either method of
accretion (a black hole would be required to avoid violating
the Eddington limit if the emission is unbeamed), but a
variable or declining flux and, perhaps, a somewhat harder
spectrum than is observed might be expected.  The presence of
a pulsar in SN 1979C has been suggested to explain the radio
light curves \citep{weiler81,pacini81}.  The behavior of very
young pulsars is unknown with a key question being the
efficiency for conversion of spin-down power to observable
x-ray emission.  X-rays would likely come mainly from a
surrounding synchrotron nebula and produce a relatively hard
power-law x-ray spectrum, mildly inconsistent with the
Chandra constraints.  With additional x-ray observations, it
would be possible to search for variability and perhaps also
to determine whether the x-ray emission is thermal line
emission or due to a continuum process and to accurately
measure the spectral parameters.

\acknowledgments  

I gratefully acknowledge the efforts of the Chandra team and
partial support from NASA grant NAG5-7405.  I thank the
referee (and editor) Greg Bothun for useful comments.  The
Digitized Sky Survey was produced at the Space Telescope
Science Institute under U.S. Government grant NAG W-2166.

%--------------

%\clearpage % uncomment for Apj, comment out for astro-ph

\end{document}